\documentclass[fleqn,usenatbib]{mnras}

\usepackage{newtxtext,newtxmath}

\usepackage[T1]{fontenc}
\usepackage{ae,aecompl} 

\usepackage{graphicx}	

\title[Chemical composition of twin-star binaries]{The chemical composition of HIP\,34407/HIP\,34426 and other twin-star comoving pairs}

\author[I. Ram\'irez et al.]{I. Ram\'irez,$^{1}$\thanks{E-mail: iramirez@tacomacc.edu}
S. Khanal,$^{2}$
S. J. Lichon,$^{1}$
J. Chanam\'e,$^{3}$
M. Endl,$^{2}$
J. Mel\'endez,$^{4}$ and
\newauthor
D. L. Lambert$^{2}$
\\
$^{1}$Tacoma Community College, 6501 South 19th Street, Tacoma, WA 98466\\
$^{2}$McDonald Observatory and Department of Astronomy, University of Texas at Austin, 2515 Speedway, Austin, TX 78712-1205\\
$^{3}$Departamento de Astronom\'ia y Astrof\'isica, Pontificia Universidad Cat\'olica de Chile, Av.\ Vicu\~na Mackenna 4860, 782-0436 Macul, Santiago, Chile \\
$^{4}$Departamento de Astronomia do IAG/USP, Universidade de S\~ao Paulo, Rua do M\~atao 1226, S\~ao Paulo, 05508-900, SP, Brasil
}

\date{Accepted 2019 September 14. Received 2019 September 13; in original form 2019 August 12}

\pubyear{2019}

\begin{document}
\label{firstpage}
\pagerange{\pageref{firstpage}--\pageref{lastpage}}
\maketitle

\begin{abstract}
We conducted a high-precision elemental abundance analysis of the twin-star comoving pair HIP\,34407/HIP\,34426. With mean error of 0.013\,dex in the differential abundances ($\Delta\mathrm{[X/H]}$), a significant difference was found: HIP\,34407 is more metal-rich than HIP\,34426. The elemental abundance differences correlate strongly with condensation temperature, with the lowest for the volatile elements like carbon around $0.05\pm0.02$\,dex, and the highest up to about $0.22\pm0.01$\,dex for the most refractory elements like aluminum. Dissimilar chemical composition for stars in twin-star comoving pairs are not uncommon, thus we compile previously-published results like ours and look for correlations between abundance differences and stellar parameters, finding no significant trends with average effective temperature, surface gravity, iron abundance, or their differences. Instead, we found a weak correlation between the absolute value of abundance difference and the projected distance between the stars in each pair that appears to be more important for elements which have a low absolute abundance. If confirmed, this correlation could be an important observational constraint for binary star system formation scenarios.
\end{abstract}


\section{Introduction}

Stars in a binary system are assumed to have equal composition because of their common origin. More precisely, their relative elemental abundances must be consistent with the relatively small range of abundances observed in individual stellar clusters and associations. However, due to the high-precision chemical analysis that is possible today, this notion has been challenged by observations \cite[e.g.,][]{ramirez11,ramirez15,teske16,tucci-maia14}. Motivated by the work of \cite{melendez09:twins}, who claimed that small abundance depletions can be produced by planet formation processes \cite[see also][]{ramirez10,ramirez14:bst}, these studies have attempted to measure with extreme precision not only overall metallicity differences in stars in binaries, but element-by-element relative abundances in {\it twin-star binaries}. The latter, binary systems in which the atmospheric parameters of the components are very similar to each other, allow us to minimize the uncertainties in a strictly differential abundance analysis, given that both stars in each binary are affected equally by systematic errors. This, in turn, is favorable to find and investigate small abundance anomalies.

High-precision chemical analyses of twin-star binaries have revealed statistically-significant differences of $\lesssim$0.05\,dex, with a couple of extreme cases where the differences amount to about 0.20\,dex,\footnote{In this work, we adopt the standard  ``bracket'' chemical abundance notation $\mathrm{[X/H]}=A(\mathrm{X}) - A(\mathrm{X})_\odot$, where $A(\mathrm{X})=\log(n_\mathrm{X}/n_\mathrm{H})+12$ and $n_\mathrm{X}$ is the number density of species X in the star's photosphere.} including the system investigated in this work. These values are comparable to the chemical abundance inhomogeneities (star-to-star scatter) found in high-precision studies ($\simeq0.01$\,dex errors in relative abundances) of stars in open clusters like the Hyades \citep{liu16}, Pleiades \citep{spina18}, and M67 \citep{liu19}. We must note, however, that strictly speaking the stellar pairs utilized in this type of binary studies are by definition comoving objects. Their reality as true, bound binary systems could be questionable and thus deserves to be further investigated.

Challenging the assumptions of chemical homogeneity of gas clouds from which systems of stars form (binaries or clusters) and unvariable surface chemical composition during the main-sequence lifetime of stars could impact severely the commendable efforts of {\it chemical tagging}. The latter attempts to reconstruct the Galaxy by grouping stars with similar composition and age, therefore assumed to have formed from the same gas cloud \citep[e.g.,][]{andrews19}. Understanding relatively small elemental abundance variations in systems expected to have common composition and figuring out where they stem from will ultimately benefit our larger-scale goals in Galactic astronomy. Furthermore, this will allow us to investigate the formation and evolution of stars and their planetary systems, and provide insights into the mixing of interstellar clouds prior to or during star and planet formation.

In this work, we present a high-precision chemical abundance analysis of the HIP\,34407/HIP\,34426 twin-star comoving pair and include our results in a compilation of 12 twin-star comoving systems (most of them likely true binaries) with high-precision chemical abundances measured. We then employ this data set to look for correlations between chemical abundance differences and stellar parameters as well as binary star system properties.

\section{The HIP\,34407/HIP\,34426 pair} \label{s:basics}

The HIP\,34407/HIP\,34426 twin-star comoving pair (HD\,54046/HD\,54100) consists of two early G-type stars with visual magnitudes around 7.7. These stars have a separation of $\simeq172$ \,arc-seconds, which at their distance of $\simeq48$\,pc corresponds to about 8\,200\,AU. With such large separation, one could reasonably expect these stars to have evolved independently from each other. Astrometric data (parallax, proper motion) and radial velocities of these stars, as measured by the \citet[][DR2]{gaia18}, are listed in Table~\ref{t:gaia_dr2}, where it is evident that while the kinematic properties of these two stars are similar, they are different enough to question whether they form a bound system. We address this issue later in this work.

\begin{table*}
\caption{Gaia DR2 data for HIP\,34407 and HIP\,34426.}
\label{t:gaia_dr2}
\centering
\begin{tabular}{lcccc}
\hline
Star & Parallax (mas) & $\mu_\alpha\cos\delta$ (mas/yr) & $\mu_\delta$ (mas/yr) & RV (km/s)\\
\hline
HIP\,34407 & $20.9719\pm0.0490$ & $-51.629\pm0.090$ & $-206.352\pm0.078$ & $-12.63\pm0.19$ \\
HIP\,34426 & $20.8673\pm0.0490$ & $-54.184\pm0.088$ & $-213.069\pm0.077$ & $-11.83\pm0.20$ \\
\hline
\end{tabular}
\end{table*}

We computed photometric temperatures for these stars using the metallicity-dependent color-temperature formulae by \cite{casagrande10}, which were calibrated using accurate effective temperatures determined with the infrared flux method. We used $\mathrm{[Fe/H]}=-0.54$ for HIP\,34426 and $\mathrm{[Fe/H]}=-0.37$ for HIP\,34407, as derived in Section~\ref{s:spectroscopy}, but note that the metallicity dependence of these photometric temperatures is very weak; errors of 0.1\,dex translate into effective temperature errors of only a few degrees. Color indices were extracted from the General Catalogue of Photometric Data \cite[][$B-V$, $b-y$]{mermilliod97} and the 2MASS \cite[][$V-J,H,K_s$; $J-K_s$]{cutri03} and {\it Hipparcos-Tycho} \cite[][$B_T-V_T$, $V_T-J,H,K_s$]{vanleeuwen97} catalogs. Excellent agreement was obtained for the temperatures derived from the ten different colors available for each star. Their weighted means and standard deviations are $T_\mathrm{eff}=6003\pm36$\,K for HIP\,34426 and $T_\mathrm{eff}=5964\pm34$\,K for HIP\,34407. Thus, their difference in photometric effective temperature is only $39\pm50$\,K (HIP\,34426$-$HIP\,34407).

Using {\it Gaia} parallaxes from their second data release \citep{gaia18} to calculate our stars' absolute magnitudes, and the photometric temperatures derived above, we computed age probability distributions based on the Yonsei-Yale stellar evolution models \citep{yi01,kim02}. These calculations were performed as in \cite{ramirez13:thin-thick}, but using the $q^2$ code \citep{ramirez14:harps} and without shifting the metallicity scale by $-0.04$\,dex, which was required for a precise relative age determination of solar twin stars (HIP\,34407 and HIP\,34426 are twins of each other, but not solar twins). As a by-product of these calculations, we also obtained ``trigonometric'' $\log\,g$ values for our stars.

Isochrone ages are sensitive to the input metallicity. Since our stars have different [Fe/H] values, it is important to carry out tests of isochrone age dependency on [Fe/H]. Moreover, in stellar evolution calculations the input metallicity corresponds to that of the stellar interior, not just the surface or convective envelope. If, as we argue later in this work, the lower [Fe/H] value of HIP\,34426 is only superficial, its internal metallicity could be assumed equal to that of HIP\,34407. Furthermore, the impact of $\alpha$-element enhancement could be significant at their relatively low [Fe/H]. We estimate an [O/Fe] ratio of about +0.13 for HIP\,34407 (Section~\ref{s:spectroscopy}), which would lead to an increase of the input metallicity for the isochrones approximately equal to $0.09$\,dex, if we utilize the scaling formula by \cite{salaris93}. This would lead to an ``internal'' [Fe/H]$=-0.28$ for both stars.

We summarize our most probable age and trigonometric $\log\,g$ results for different input [Fe/H] values in Table~\ref{t:age}. In all cases the uncertainties in age and $\log\,g$ are the same: 0.8\,Gyr and 0.02, respectively. When we utilize the surface [Fe/H] values obtained in this work, there is an age difference of almost 1 Gyr between the stars. The age difference reduces to about 0.4\,Gyr when we adopt an equal value of [Fe/H]$=-0.37$ for both stars, while they are almost identical when [Fe/H]=$-0.28$ is adopted instead. As a higher metallicity is adopted, the derived age becomes younger. Thus, when taking into consideration our assumption that the lower [Fe/H] of HIP\,34426 is due to metal depletion of the convective envelope only and the effect of the small, but not negligible $\alpha$-element enhancement, the isochrone ages of these two stars become more consistent with each other. Moreover, the trigonometric $\log\,g$ values approach those derived from spectroscopy alone when the common, higher [Fe/H] is adopted, even though it seems somewhat low for HIP\,34426.

\begin{figure}
\includegraphics[width=\columnwidth, trim=0 1.5cm 0 0.2cm]{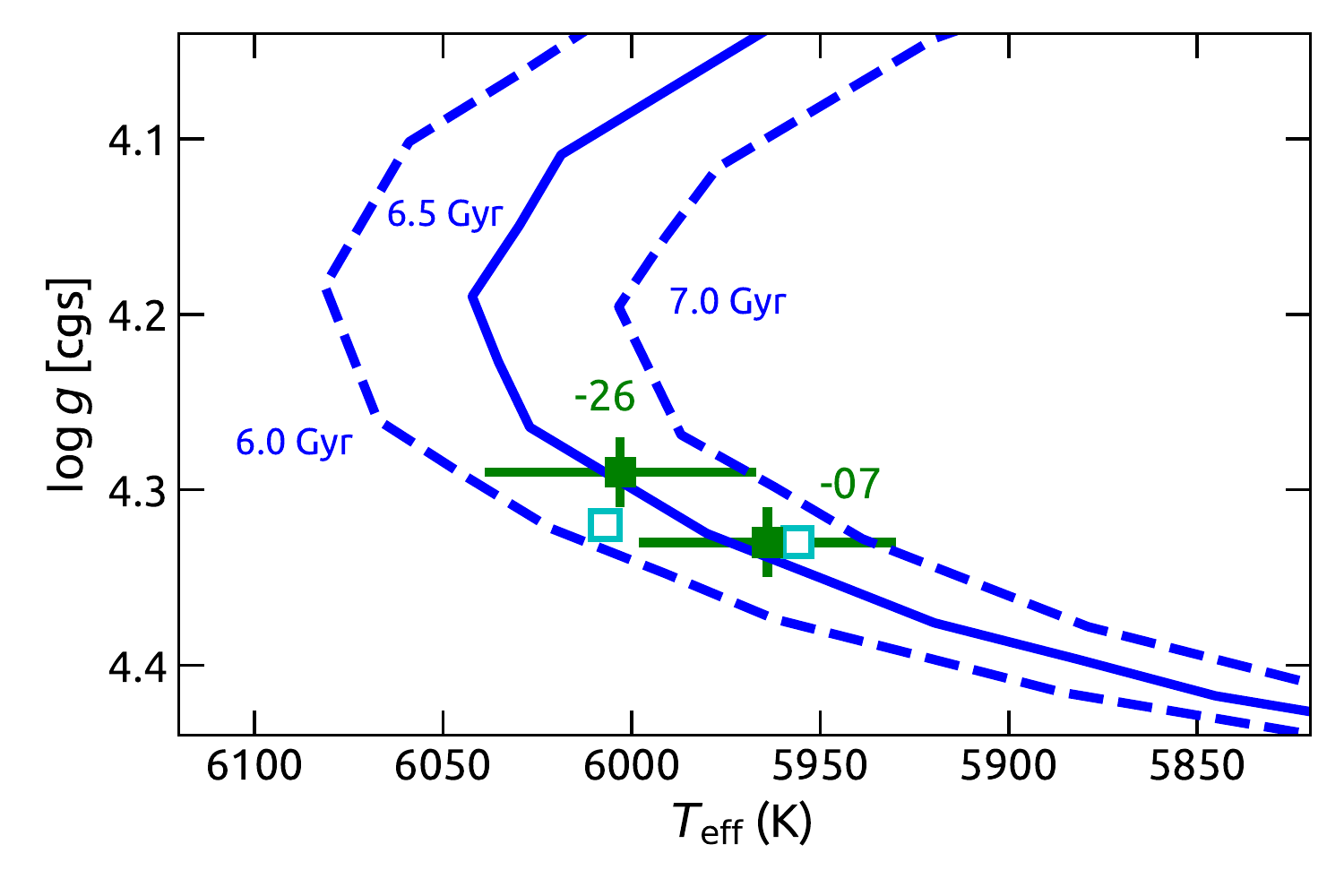}
\caption{Effective temperature versus surface gravity diagram. The blue solid line corresponds to a 6.5\,Gyr isochrone, while the blue dashed lines represent isochrones of 6.0 Gyr and 7.0 Gyr of age. The metallicity adopted for these isochrones is $\mathrm{[Fe/H]}=-0.28$, which corresponds to a common, $\alpha$-element enhanced internal metallicity. The locations of HIP\,34407 (labeled ``-07'' in the diagram) and HIP\,34426 (labeled ``-26'' in the diagram), as determined using color $T_\mathrm{eff}$ and trigonometric $\log\,g$ values are indicated with the green filled squares with  error bars. The cyan open squares correspond to the locations of these stars when the purely spectroscopic stellar parameters are adopted instead.}
\label{f:hr}
\end{figure}

With an $\alpha$-element enhanced, equal internal input [Fe/H]=$-0.28$, we obtained the most consistent ages of $6.5\pm0.8$\,Gyr for HIP\,34426 and $6.6\pm0.8$\,Gyr for HIP\,34407. The locations of these stars on the $T_\mathrm{eff}$ versus $\log\,g$ diagram is shown in Figure~\ref{f:hr}, along with theoretical isochrones of 6.0, 6.5, and 7.0~Gyr.

The relatively low [Fe/H] and old age values obtained for this pair compared to the solar neighborhood distributions \cite[e.g.,][]{allende04:s4n} suggest that a coincidental alignment of these two comoving stars is unlikely. Their relatively large separation, on the other hand, argues against the reality of this pair as a true binary system. We acknowledge this as a potential limitation of our analysis and attempt to address it in a later section where we investigate statistical properties of a sample of wide binary candidates that includes this system (Section~\ref{s:compilation}).

\begin{table}
\caption{Isochrone age and trigonometric $\log\,g$ values obtained for different choices of the input [Fe/H].}
\label{t:age}
\centering
\begin{tabular}{lccc} \hline
Star & [Fe/H] & Age (Gyr) & $\log\,g$ \\ \hline
HIP\,34407 & $-0.37$ & 7.8 & 4.30 \\
HIP\,34426 & $-0.54$ & 8.7 & 4.25 \\ \hline
HIP\,34407 & $-0.37$ & 7.8 & 4.30 \\
HIP\,34426 & $-0.37$ & 7.4 & 4.27 \\ \hline
HIP\,34407 & $-0.28$ & 6.6 & 4.33 \\
HIP\,34426 & $-0.28$ & 6.5 & 4.29 \\ \hline
\end{tabular}
\end{table}

\section{Spectroscopic data and chemical analysis} \label{s:spectroscopy}

High-resolution ($R\simeq65\,000$), high signal-to-noise ratio ($S/N\simeq500$ per pixel) spectra of HIP\,34407, HIP\,344026, and a bright asteroid (reflected sunlight from asteroid Vesta) were acquired with the MIKE spectrograph on the 6.5\,m Magellan Clay Telescope on 23 January 2012. The wavelength coverage goes from 3960 to 8290\,\AA. We used exposure times of 30 minutes for HIP\,34407 and 44 minutes for HIP\,34426. Data reduction was done using the MIKE Carnegie-Python pipeline. Continuum normalization and merging of orders were completed in the standard manner using the IRAF's tasks {\it scopy}, {\it continuum}, and {\it scombine}.

\begin{figure}
\includegraphics[width=\columnwidth, trim=0cm 1.5cm 0 0.1cm]{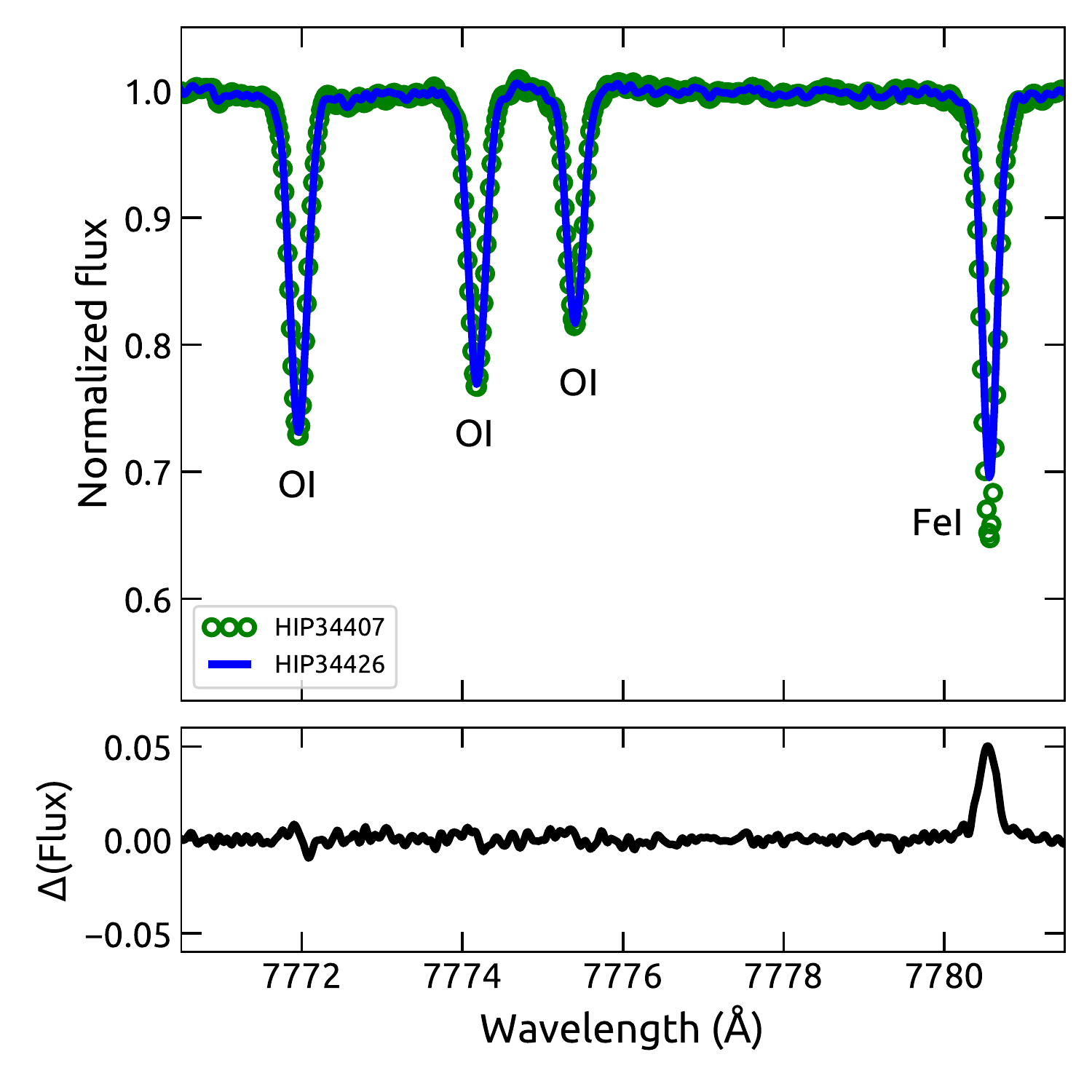}
\caption{Top panel: Spectra of HIP\,34407 (green open circles) and HIP\,34426 (blue solid line) around the 777\,nm oxygen triplet region. Bottom panel: Normalized flux difference between HIP\,34407 and HIP\,34426 in the same wavelength region.}
\label{f:oxygen_triplet}
\end{figure}

A small portion of the HIP\,34407 and HIP\,34426 spectra after data reduction can be seen in Figure~\ref{f:oxygen_triplet}. It is apparent in this figure that the oxygen lines are of almost the same equivalent width whereas the iron line shows a clear difference between the two stars. Such obvious difference in volatiles (like oxygen) versus refractories (like iron) in a twin-star binary candidate is outstanding among other similar systems investigated before. The wavelength region between 7776.5\ and 7779.5\,\AA\ is free from absorption lines and could therefore be employed to estimate the $S/N$ of these spectra. The inverse of the standard deviation of the $\Delta$(Flux) values in this region is 594, which implies a $S/N=840$ for each spectrum. The counts in the MIKE data are highest in this region of the spectrum, thus it is not surprising that the estimated $S/N$ is higher than the average value of about 500.

The atmospheric parameters $T_\mathrm{eff}$, $\log\,g$, [Fe/H], and $v_t$ (microturbulent velocity) were measured using a standard iron line analysis. The iron line equivalent widths, as all the other elements, were measured using the {\it splot} task in IRAF. As described in detail in Sect.\ 4.1 of \cite{ramirez15}, the MOOG code by \cite{sneden73} was used to perform the translations from line strength to abundances using the curve-of-growth method, while the $q^2$ Python package (essentially a MOOG wrapper) was employed to derive the atmospheric parameters semi-automatically.\footnote{The Qoyllur-Quipu ($q^2$) code and a user tutorial are available online at \url{https://github.com/astroChasqui/q2.}}

\begin{figure}
\includegraphics[width=\columnwidth, trim=0 1cm 0 0]{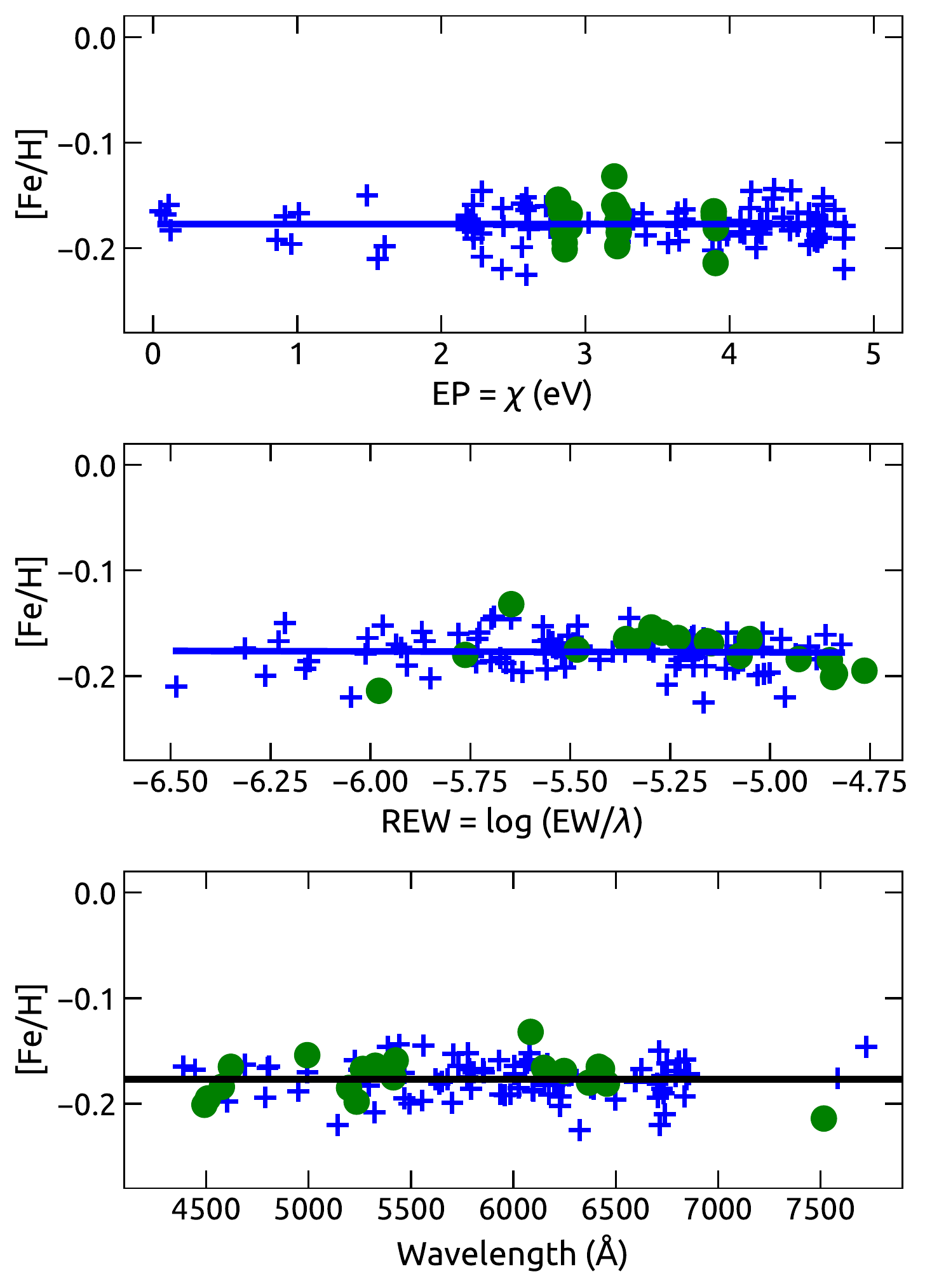}
\caption{Line-by-line iron abundance of HIP\,34426, measured relative to HIP\,34407, as a function of excitation potential (top panel), reduced equivalent width (middle panel), and wavelength (bottom panel). Blue crosses represent \ion{Fe}{i} lines and filled green circles correspond to \ion{Fe}{ii} lines. The blue solid lines on the top and middle panels are linear fits to the \ion{Fe}{i} data while the solid black line in the lower panel corresponds to the average iron abundance from all lines.}
\label{f:q2}
\end{figure}

Initially, we conducted an analysis of each star relative to the Sun, adopting as solar parameters $T_\mathrm{eff}=5777$\,K, $\log\,g=4.44$, [Fe/H]$=0$, and $v_t=1.0$\,km/s. For all \ion{Fe}{i} and \ion{Fe}{ii} features, line-by-line differential abundances were measured. The differential iron abundance is compared to the excitation potential ($\chi$) and reduced equivalent width (REW) of the lines. The atmospheric parameters were fine-tuned such that there would be no correlation (zero slope) in the [Fe/H] versus $\chi$ and [Fe/H] versus REW graphs. To improve on the precision, the process was repeated, but using the star HIP\,34407 as reference (and their parameters as determined in the previous solar-reference analysis). The final result (HIP\,34426 minus HIP\,34407) is shown in Figure~\ref{f:q2}, where it can be clearly seen that HIP\,34426 has an average iron abundance about $0.18\pm0.01$\,dex lower than HIP\,34407.

\begin{figure}
\includegraphics[width=\columnwidth, trim=0.1cm 1.5cm 0 0.4cm]{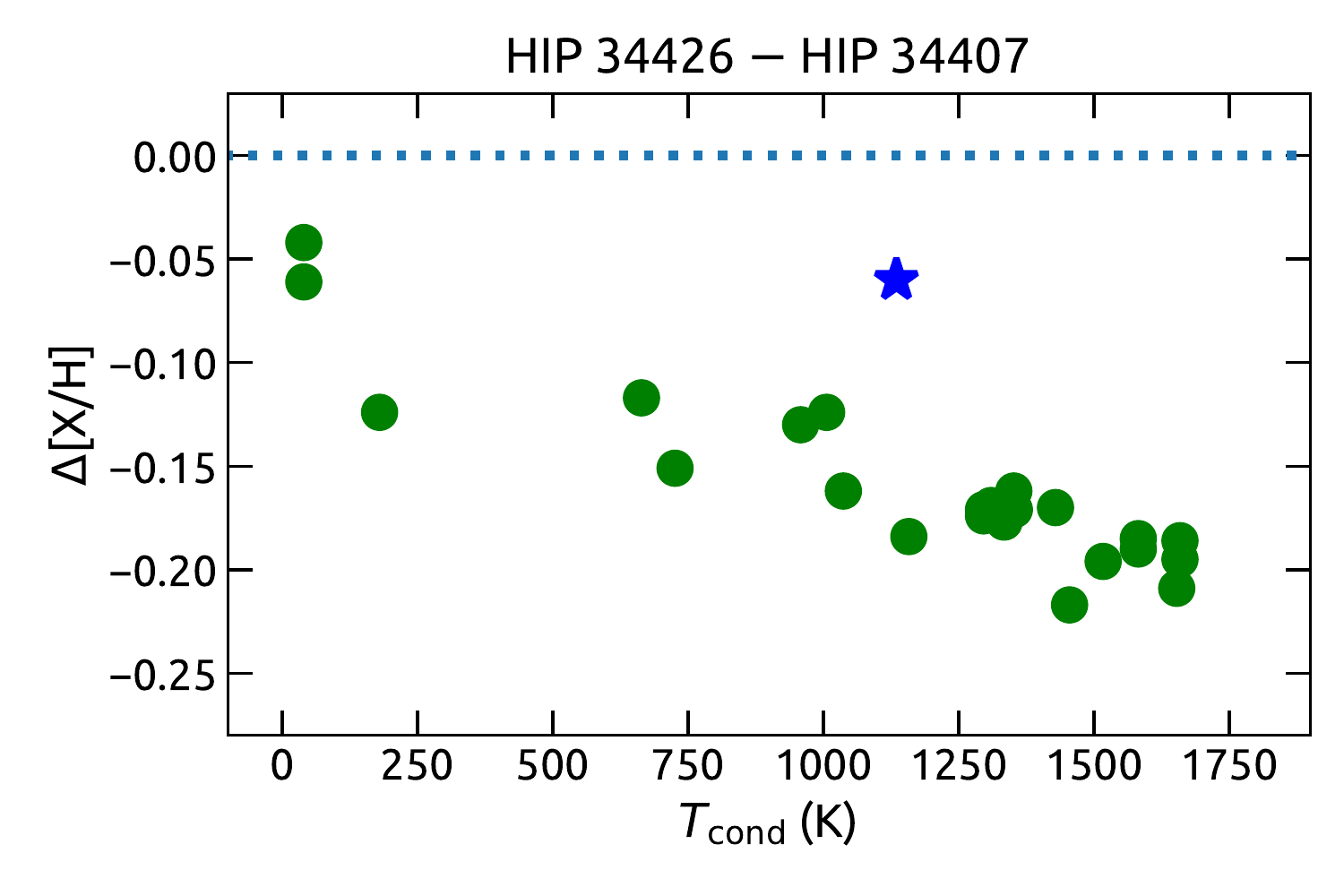}
\caption{Elemental abundance differences between HIP\,34426 and HIP\,34407 as a function of the elements' condensation temperature. The blue star symbol corresponds to lithium.}
\label{f:tc}
\end{figure}

For consistency with previous work on twin-star binaries, we used the spectroscopically-derived $\log\,g$ values in the determination of detailed chemical abundances. The abundances of 22 elements were measured for this star system using the line-list by \citet[][their Table~3]{ramirez15}. The carbon abundance was acquired using \ion{C}{i} and CH lines while for Ti, Cr, and Fe, both neutral and singly-ionized species' lines were employed. For the oxygen abundance, the oxygen triplet at 777\,nm was used, adopting the non-LTE corrections by \cite{ramirez07}. The lithium abundances were measured as in \cite{ramirez11} using spectrum line fits with the synth driver in MOOG. Non-LTE corrections for the lithium abundances were applied following \cite{lind09}. Our results for stellar parameters and elemental abundance are listed in Tables~\ref{t:parameters} and \ref{t:absolute_abundances}. The more precise differential abundances are given in Table~\ref{t:abundances}. On an element-to-element basis, the magnitude of the differential abundance increases with higher condensation temperature,\footnote{The $T_\mathrm{cond}$ values adopted in this work correspond to the 50\,\%\ $T_\mathrm{cond}$ values computed by \cite{lodders03} for a solar-composition gas.} as seen in Figure~\ref{f:tc}, a trend that all elements, except lithium, follow tightly.

\begin{table}
\caption{Stellar parameters for the HIP\,34407/HIP\,34426 binary system.}
\label{t:parameters}
\centering
\begin{tabular}{lccc}
\hline
Star & $T_\mathrm{eff}$ (K) & $\log\,g$ [cgs] & [Fe/H] (dex) \\
\hline
HIP\,34407 & $5956\pm12$ & $4.33\pm0.03$ & $-0.367\pm0.010$ \\
HIP\,34426 & $6007\pm8$ & $4.32\pm0.02$ & $-0.544\pm0.006$ \\
\hline
\end{tabular}
\end{table}

\begin{table}
\caption{Absolute abundances, $A_\mathrm{X}$ [dex], in HIP\,34407 and HIP\,34426.}
\label{t:absolute_abundances}
\centering
\begin{tabular}{lcc}
\hline
Species & HIP\,34407 & HIP\,34426 \\ \hline
LiI & $2.372\pm0.050$ & $2.312\pm0.050$ \\
CI & $8.062\pm0.082$ & $8.020\pm0.067$ \\
CH & $7.814\pm0.045$ & $7.753\pm0.022$ \\
OI & $8.448\pm0.016$ & $8.324\pm0.013$ \\
NaI & $5.869\pm0.034$ & $5.739\pm0.044$ \\
MgI & $7.304\pm0.062$ & $7.136\pm0.051$ \\
AlI & $6.111\pm0.068$ & $5.902\pm0.068$ \\
SiI & $7.235\pm0.058$ & $7.065\pm0.065$ \\
SI & $6.699\pm0.116$ & $6.582\pm0.165$ \\
KI & $5.130\pm0.100$ & $5.006\pm0.100$ \\
CaI & $6.066\pm0.070$ & $5.870\pm0.066$ \\
ScII & $2.876\pm0.126$ & $2.689\pm0.132$ \\
TiI & $4.623\pm0.067$ & $4.438\pm0.072$ \\
TiII & $4.672\pm0.105$ & $4.482\pm0.113$ \\
VI & $3.594\pm0.132$ & $3.423\pm0.127$ \\
CrI & $5.219\pm0.074$ & $5.045\pm0.074$ \\
CrII & $5.220\pm0.071$ & $5.049\pm0.071$ \\
MnI & $4.789\pm0.122$ & $4.605\pm0.124$ \\
FeI & $7.085\pm0.066$ & $6.908\pm0.069$ \\
FeII & $7.076\pm0.067$ & $6.901\pm0.076$ \\
CoI & $4.575\pm0.108$ & $4.413\pm0.129$ \\
NiI & $5.819\pm0.078$ & $5.648\pm0.098$ \\
CuI & $3.702\pm0.083$ & $3.539\pm0.068$ \\
ZnI & $4.121\pm0.058$ & $3.971\pm0.064$ \\
YII & $1.672\pm0.025$ & $1.478\pm0.028$ \\
BaII & $1.784\pm0.033$ & $1.567\pm0.038$ \\
\hline
\end{tabular}
\end{table}

\begin{table}
\caption{Elemental abundance differences between HIP\,34426 and HIP\,34407.}
\label{t:abundances}
\centering
\begin{tabular}{lccc}
\hline
Species & $T_\mathrm{cond}$ (K) & $\Delta$[X/H] (dex) & error (dex) \\
\hline
LiI  & 1135 & $-0.050$ & 0.040 \\
CI   &   40 & $-0.042$ & 0.019 \\
CH   &   40 & $-0.061$ & 0.019 \\
OI   &  180 & $-0.124$ & 0.009 \\
NaI  &  958 & $-0.130$ & 0.010 \\
MgI  & 1336 & $-0.168$ & 0.012 \\
AlI  & 1653 & $-0.209$ & 0.007 \\
SiI  & 1310 & $-0.169$ & 0.005 \\
SI   &  664 & $-0.117$ & 0.027 \\
KI   & 1006 & $-0.124$ & 0.011 \\
CaI  & 1517 & $-0.196$ & 0.007 \\
ScII & 1659 & $-0.186$ & 0.013 \\
TiI  & 1582 & $-0.185$ & 0.008 \\
TiII & 1582 & $-0.190$ & 0.009 \\
VI   & 1429 & $-0.170$ & 0.014 \\
CrI  & 1296 & $-0.174$ & 0.009 \\
CrII & 1296 & $-0.171$ & 0.012 \\
MnI  & 1158 & $-0.184$ & 0.008 \\
FeI  & 1334 & $-0.177$ & 0.006 \\
FeII & 1334 & $-0.175$ & 0.009 \\
CoI  & 1352 & $-0.162$ & 0.017 \\
NiI  & 1353 & $-0.171$ & 0.007 \\
CuI  & 1037 & $-0.162$ & 0.013 \\
ZnI  &  726 & $-0.151$ & 0.009 \\
YII  & 1659 & $-0.195$ & 0.009 \\
BaII & 1455 & $-0.217$ & 0.009 \\
\hline
\end{tabular}
\end{table}

Unlike other elements, it is very well-known that lithium gets depleted on the surfaces of sunlike stars over time \cite[e.g.,][]{carlos16,carlos19}. This is because lithium burns at the temperatures found close to the base of the convection zone in this type of stars. Although not precisely constrained, the statistical analysis of over 600 solar-type stars by \cite{ramirez12_lithium} showed that lithium depletion is enhanced with lower effective temperature and higher metallicity around the solar parameters. Given that HIP\,34407 is cooler and more metal rich than HIP\,34426, one would expect the present-day lithium abundance of HIP\,34407 to be more depleted than that of HIP\,34426. While both had their [Li/H] reduced, that of HIP 34407 became an even smaller number, therefore increasing the differential $\Delta$[Li/H] value (in the HIP\,34426 minus HIP\,34407 sense). Indeed, Figure~\ref{f:tc} shows that $\Delta$[Li/H] (shown with the star symbol) is too high relative to the clear trend that the rest of elements follow.

The iron abundance difference observed in most twin-star binaries analyzed with high precision to date is about 0.05\,dex or lower. A notable exception is the case of the HD\,240430/HD\,240429 pair, which revealed a difference of nearly 0.20\,dex in the study by \cite{oh18}. In HIP\,34426, the iron abundance is also about 0.20\,dex lower than in its binary companion, HIP\,34407. In fact, the smallest difference in elemental abundance for the HIP\,34426/HIP\,34407 pair is about 0.05\,dex, for carbon. The most refractory elements like aluminum and scandium show abundance differences closer to 0.20\,dex. The $\Delta$[X/H] versus $T_\mathrm{cond}$ plot for the ``Kronos \& Krios'' pair studied by \citet[][their Figure~8]{oh18}, which is a system with a similarly large [Fe/H] difference, looks remarkably similar to our Figure~\ref{f:tc}, although they find very small differences for Na and Mn (about 0.02\,dex), which appear to break the trend. In the HIP\,34426/HIP\,34407 system, the abundance differences appear to correlate with $T_\mathrm{cond}$ much more smoothly.

The old age and relatively low metallicity of the HIP\,34407/HIP\,34426 pair naturally brings into question its Galactic population membership: thin disk or thick disk. The Galactic space velocities of these stars are $U,V,W\simeq20,-30,-30$\,km/s \cite[e.g.,][]{casagrande11}, which places the system well within the thin-disk kinematic distribution on the Toomre $V$ versus $\sqrt{U^2+W^2}$ diagram \cite[e.g.,][their Figure 3]{ramirez07}. A quick inspection of the system's $\alpha$-element abundances appear to confirm its thin-disk membership. For example, the [Mg/Fe] ratio of HIP\,34407 is about 0.07, its [O/Fe] ratio is 0.13, while the [Ti/Fe] ratio of HIP\,34427 is approximately 0.03.\footnote{For these estimates, we employed the solar abundances by \cite{Asplund09:review} and our measured values from Table~\ref{t:abundances}.} These values place the stars on the lower end of the [X/Fe] versus [Fe/H] plots typically utilized to disentangle the Galactic disk stellar populations \cite[e.g.,][]{reddy06,ramirez07,bensby14}. Overall, the $\alpha$-element abundances of both HIP\,34407 and HIP\,34426 are low, which characterizes thin-disk stars at relatively low metallicity.

\section{The Condensation Temperature Trend}

Condensation temperature trends like the one we have discovered in the HIP\,34426/HIP\,34407 system have been attributed to planet formation processes. In particular, it has been argued that the formation of rocky planets early in the history of a planetary system takes away from the star high condensation temperature elements more than low condensation temperature elements. Thus, in a binary system, one would expect the star hosting more rocky planets to be slightly poor in refractory elements.

\begin{figure}
\includegraphics[width=\columnwidth, trim=0 1.3cm 0 0.2cm]{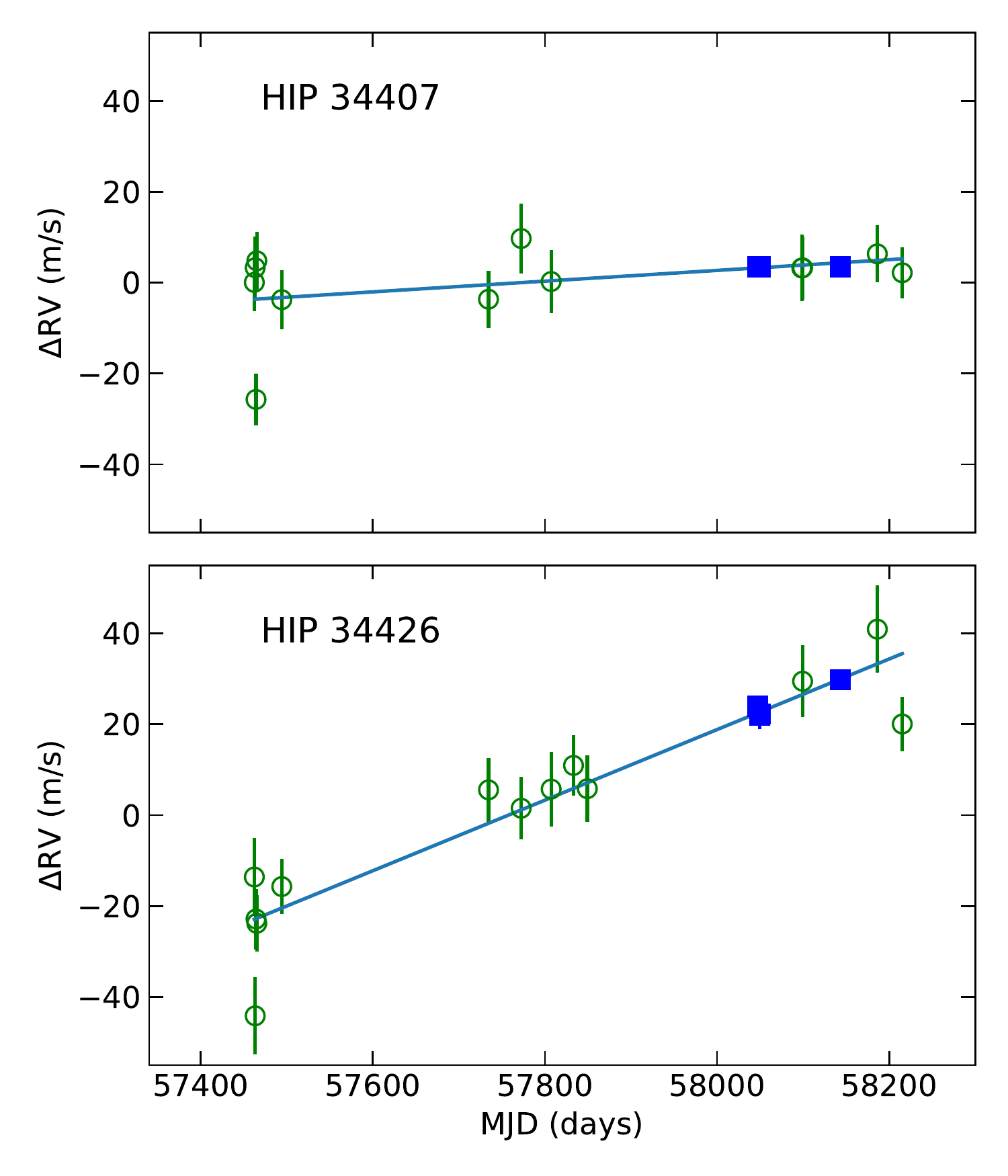}
\caption{Radial velocity data for HIP\,34407 (upper panel) and HIP\,34426 (lower panel). The green open circles correspond to measurements made with the Tull spectrograph at McDonald Observatory while the filled blue squares represent measurements by the HARPS spectrograph. The solid lines are linear fits to the data, which have been arbitrarily shifted to have an average value of zero.}
\label{f:rv}
\end{figure}

Following this logic, one would expect HIP\,34426 to be the star with more rocky planets formed around it. We do not have enough data yet to confirm or reject the presence of small planets around either one of these stars, but we have started a radial velocity follow-up of the system using the Tull spectrograph at McDonald Observatory as well as the HARPS spectrograph on the 3.6-m Telescope at La Silla. The data collected at these two facilities, as of this writing, are shown in Figure~\ref{f:rv}.

While HIP\,34407 appears to have nearly constant radial velocity, its companion, HIP\,34426, the metal-poor component of the system, shows a clear linear trend. We have observed at least a 60\,m/s variation in the radial velocity of HIP\,34426, suggesting either the presence of a large Jupiter-like planet or possibly a third star member of the system in orbit around HIP\,34426.

An alternative to the metal depletion scenario described above is the idea that planetary material, presumably rich in refractories, would have been ingested by one of the stars following the migration of large planets. In this case, the more metal-rich star would be the one that engulfed its planets. One would expect in this case also a higher level of lithium content relative to the companion, particularly given how old this system is. If HIP\,34407 has accreted planets, its lithium abundance should be higher than that of HIP\,34426. What we observe is the opposite. The lithium content of HIP\,34426 is higher relative to the $T_\mathrm{cond}$ trend. As explained in Section~\ref{s:spectroscopy}, this could be reasonably explained by the small differences in stellar parameters of the two components of this binary. Most likely, explaining the lithium content of these two stars does not require an extreme pollution of one of them due to planetary ingestion. The accretion of a planet or brown dwarf would also lead to an increased rate of rotation in the star on which this occurred. However, the projected rotational velocities ($V\sin i$) of these stars are very similar, as evidenced by Figure~\ref{f:oxygen_triplet}.

The radial velocity data is of course inconclusive with regards to the presence or absence of small, Earth-like planets, but this is also the case for all other studies in this particular field. It is important, however, to continue investigating the planet populations around these stars. Only with large statistics and a variety of systems investigated in this manner we will be able to begin disentangling the different scenarios.

Abundance trends with condensation temperature that resemble the one we observe in the HIP\,34407/HIP\,34426 system have been seen also in RV\,Tau stars by, for example, \cite{giridhar05}, although in these stars the effect is orders of magnitude stronger. One interpretation of these trends with $T_\mathrm{cond}$ involves dust-gas separation which requires a site for dust to form, for example a circumstellar disk.

\begin{figure}
\includegraphics[width=\columnwidth, trim=0 1.5cm 0 0]{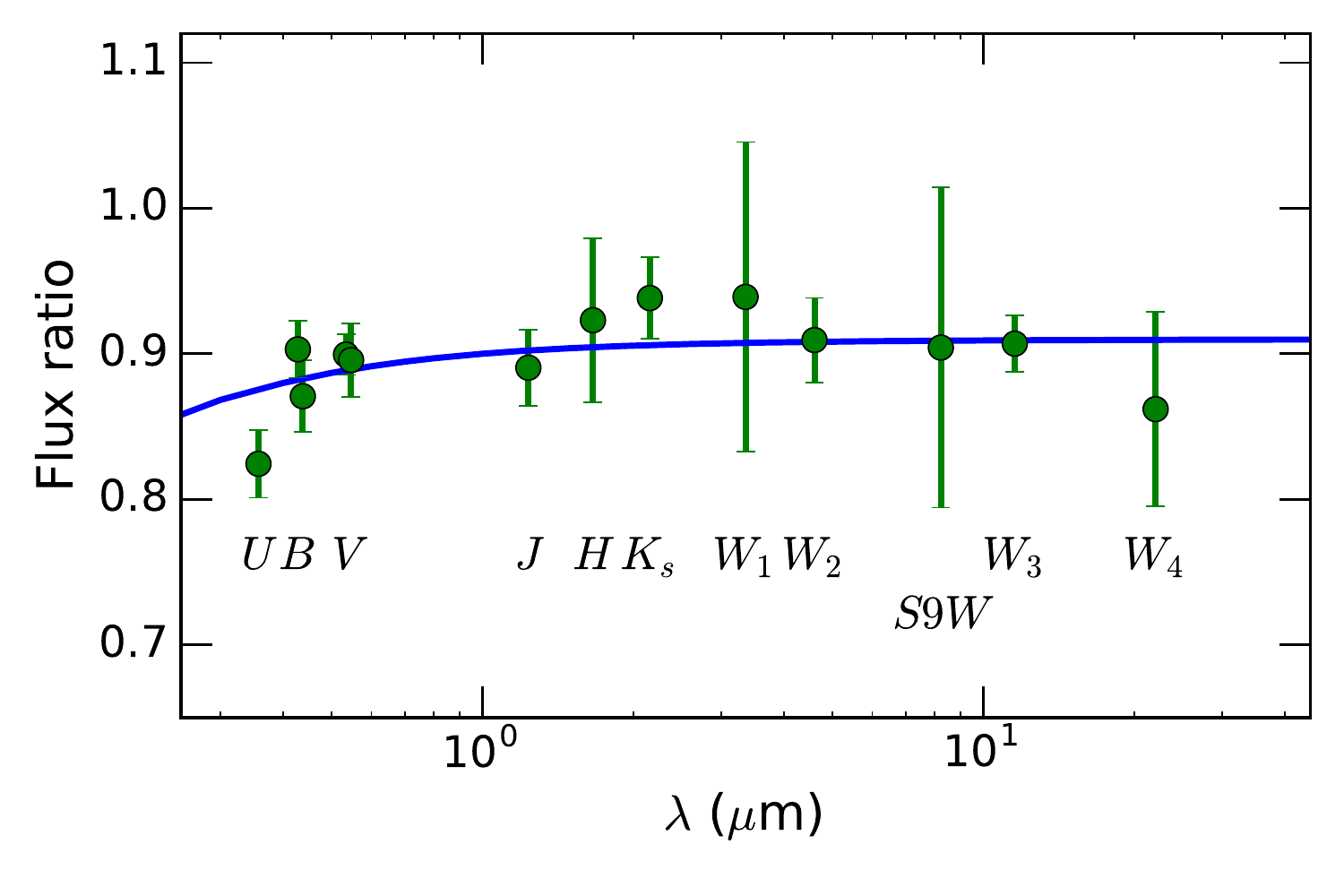}
\caption{Ratio of wavelength-dependent fluxes (HIP\,34407/HIP\,34426) determined from photometric data using the VOSA analyzer.}
\label{f:photometry}
\end{figure}

In order to investigate the possibility of circumstellar material around the HIP\,34407/HIP\,34426 pair, we searched for as much reliable photometric data as possible using the Virtual Observatory SED (Spectral Energy Distribution) Analyzer (VOSA, Bayo et al. 2008).\footnote{\url{http://svo2.cab.inta-csic.es/theory/vosa/}} Both of these stars have $U,B,V$ magnitudes in the \cite{mermilliod94} compilation, as well as 2MASS \citep{skrutskie06} and WISE \citep{wright10} photometry, in addition to magnitudes from one of the AKARI Mission bands \citep{murakami07}. We employed VOSA to calculate physical fluxes and to search for infrared excess in each star's SED. No excess was detected in either one of the two stars according to the VOSA algorithm, which is a refined version of the classic \cite{lada06} ``$\alpha$-parameter'' criterion. A plot of the flux ratio (HIP\,34407/HIP\,34426) as a function of wavelength ($\lambda$) is given in Figure~\ref{f:photometry}, which exhibits no significant $\lambda$-dependence. The fact that the flux ratio is about 0.9 can be fully explained by the difference in the stars' sizes. From our isochrone analysis (Section~\ref{s:basics}), we estimate radii of 1.11 and 1.16\,$R_\odot$ for HIP\,34407 and HIP\,34426, respectively. The ratio of these radii squared, which is proportional to the flux ratio, is 0.92. The average flux ratio is somewhat lower than this value due to the slightly lower effective temperature of HIP\,34407. Indeed, the solid line in Figure~\ref{f:photometry} shows the ratio of two blackbodies with temperatures that correspond to HIP\,34407 and HIP\,34426, scaled by the ratio of the stars' radii squared. With the exception of the $U$ band, and marginally the $K_s$ band, this simple model matches the data within the errors, suggesting no differential infrared excess, and therefore no clear evidence of circumstellar dust around either one of the HIP\,34407/HIP\,34426 stars or at least no significant difference between them that could explain the observed elemental abundance differences as due to dust-gas separation.

We have also looked into recent studies on the effects of difussion on photospheric abundances, particularly those attempting to explain the observational data of the M67 open cluster \citep{bertelli18,gao18,souto18,souto19}. The abundance differences we observe in the HIP\,34407/HIP\,34426 pair are comparable to the range of abundances seen in the M67 cluster, which are roughly consistent with difussion model predictions. The HIP\,34407/HIP\,34426 pair is about twice as old as M67, thus one would expect even higher difussion effects.

Difusion models predict a strong dependence on $\log\,g$, with the largest effects occuring around $\log\,g=4.2$, which is not too far off from our stars' surface gravity values. Naturally, difussion alone could only explain the trends we observe if it produces an abundance difference over time. The stars being nearly twins of each other minimizes any potential systematic effects like that of diffusion. In fact, since both our stars have basically the same $\log\,g$, one would expect these differential difusion effects to be negligible, unless turbulence is very different between these two stars' interiors, which also seems unlikely.

\section{Other twin-star binaries} \label{s:compilation}

\begin{table*}
\caption{Twin-star binary systems with high-precision chemical abundances previously measured.}
\label{t:binaries}
\centering
\begin{tabular}{lcrccrrccl}
\hline
System & $T_\mathrm{eff,avg}$ (K) & $|\Delta T_\mathrm{eff}|$ & $\log\,g_\mathrm{avg}$ & $|\Delta\log\,g|$ & Sep. (AU) & $\mathrm{[Fe/H]}_\mathrm{avg}$ & $|\Delta\mathrm{[Fe/H]}|$ & $\sigma(\Delta\mathrm{[Fe/H]})$ & Source \\
\hline
XO-2	& 5470 & 60 & 4.44 & 0.02 & 4500 & 0.374 & 0.054 & 0.005 & \cite{ramirez15} \\
16 Cygni & 5791 & 79 & 4.33 & 0.05 & 860 & 0.078 & 0.047 & 0.005 & \cite{tucci-maia14} \\
WASP-94 & 6153 & 82 & 4.23 & 0.09 & 2700 & 0.313 & 0.015 & 0.004 & \cite{teske16} \\
Kronos \& Krios & 5841 & 75 & 4.38 & 0.10 & 11704 & 0.105 & 0.190 & 0.010 & \cite{oh18} \\
HAT-P-1 & 6150 & 202 & 4.40 & 0.07 & 1550 & 0.151 & 0.009 & 0.009 & \cite{liu14} \\
HAT-P-4 & 6037 & 1 & 4.36 & 0.05 & 28446 & 0.226 & 0.105 & 0.006 & \cite{saffe17} \\
$\zeta^2$ Reticuli & 5782 & 144 & 4.54 & 0.01 & 3713 & -0.205 & 0.020 & 0.003 & \cite{saffe16} \\
HD 20781-82 & 5557 & 465 & 4.46 & 0.10 & 9000 & 0.010 & 0.060 & 0.010 & \cite{mack14} \\
HD 80606-07 & 5587 & 52 & 4.45 & 0.04 & 1200 & 0.350 & 0.000 & 0.040 & \cite{mack16} \\
HIP 99727-29 & 5751 & 90 & 4.18 & 0.04 & 2761 & 0.191 & 0.011 & 0.018 & \cite{ramirez14:bst} \\
HD 134439-40 & 5015 & 98 & 4.67 & 0.02 & 8617 & -1.410 & 0.040 & 0.028 & \cite{reggiani18} \\
HIP 34407-26 & 5984 & 39 & 4.27 & 0.06 & 8200 & -0.445 & 0.177 & 0.006 & This Work \\
\hline
\end{tabular}
\end{table*}

To the best of our knowledge, there have been 11 other high-precision chemical abundance studies of (candidate) binary star systems that include solar twins and analogs, or twin stars of the F and G spectral type. In most of these studies, a statistically significant difference in elemental abundances has been reported. We compiled the results from all these studies to search for correlations with stellar parameters and binary star properties. The list of systems analyzed in this work is given in Table~\ref{t:binaries}. The complete dataset, including detailed abundance values, is available upon request. To calculate the separation values, we used data from Gaia’s second data release \citep{gaia18} along with the \textit{separation\_3d} and \textit{SkyCoord} functions of astropy \citep{astropy:2013,astropy:2018}. We took the mean of the parallax values for each of the binary star systems to calculate their average distances. Thus, strictly speaking, the separation values used in this work are accurate representations of the \textit{projected} binary separations.

\begin{figure}
\includegraphics[width=\columnwidth, trim=1.7cm 7.5cm 1.6cm 10cm]{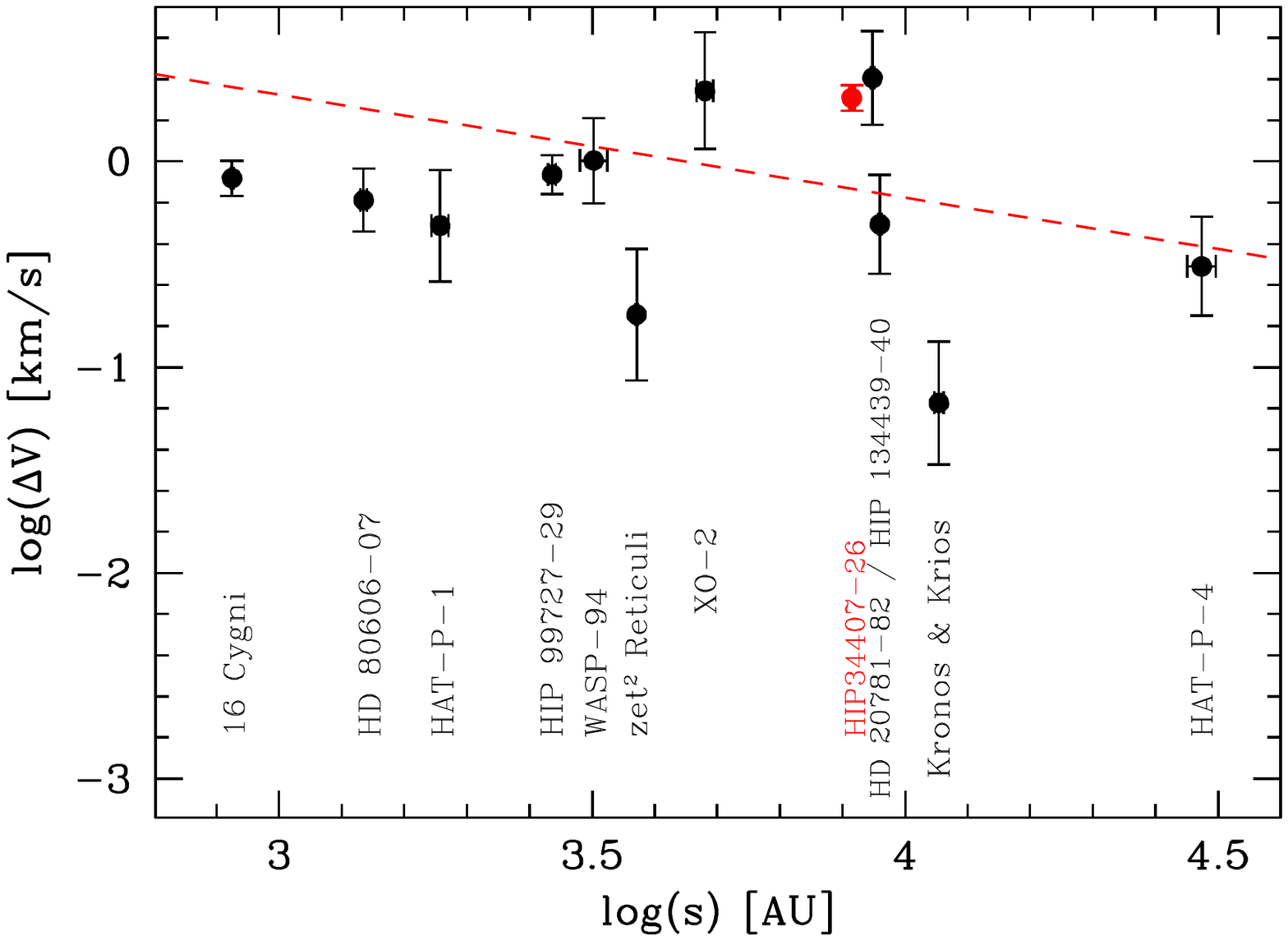}
\caption{Log-log plot of projected separation versus total velocity difference for the 12 twin-star binary systems listed in Table~\ref{t:binaries}. The red dashed line corresponds to bound pairs with total mass $10\,M_\odot$ in circular orbits, which represents a reliable boundary for true binarity.}
\label{f:binarity}
\end{figure}

Following \cite{andrews17}, we examined the location of our binary star candidates on the $\log(s)-\log(\Delta\,V)$ plot, shown in Figure~\ref{f:binarity}, to determine whether they are currently bound systems. The $s$ in this diagram is the projected separation in AU and $\Delta V$ represents the difference in total velocity in km/s. The dashed line in Figure~\ref{f:binarity} corresponds to bound binaries with total mass of $10\,M_\odot$ in circular orbits. As argued by \cite{andrews17}, pairs that fall below that line in the plot are considered physically bound. For each star in our twin-star binary system candidate sample, a 1-$\sigma$ error bar is shown, which was created by accounting for uncertainties in the parallaxes, radial velocities, and proper motions of the stars in each pair. We note that the criterion employed in this work to evaluate whether a comoving pair is bound utilizes total velocity, not just radial velocity, as projection effects might not be negligible in this context, particularly at large separations \citep{shaya11,el-badry19}

Assuming there are no unknown systematics in the Gaia DR2 data at the level shown by the vertical axis, Figure~\ref{f:binarity} would suggest that the HIP\,34407/HIP\,34426 and HD\,134439/HD\,13440 pairs are either not bound systems (while XO-2 would seem marginally bound), or else some other effect is inflating the measured relative velocity between their components, like an unseen companion to any of the components would do. Interestingly, these pairs appear to be some of the oldest systems in the sample. Thus, perhaps their non-bound nature can be explained by secular disruption due to encounters with Galactic substructures as well as large-scale tidal field effects. Their reality as true binary systems (at formation) might not be too questionable after all. Chance alignment with such similar composition in a Galactic chemical evolution sense is unlikely, while systematics in the astrometry might be significant. The analysis below assumes that all pairs in Table~\ref{t:binaries} formed as true binary systems. Later in this section, we evaluate the effect of excluding the two potentially unbound systems from our analysis.

\begin{figure}
\includegraphics[width=\columnwidth, trim=0 1.7cm 1.2cm 1.5cm]{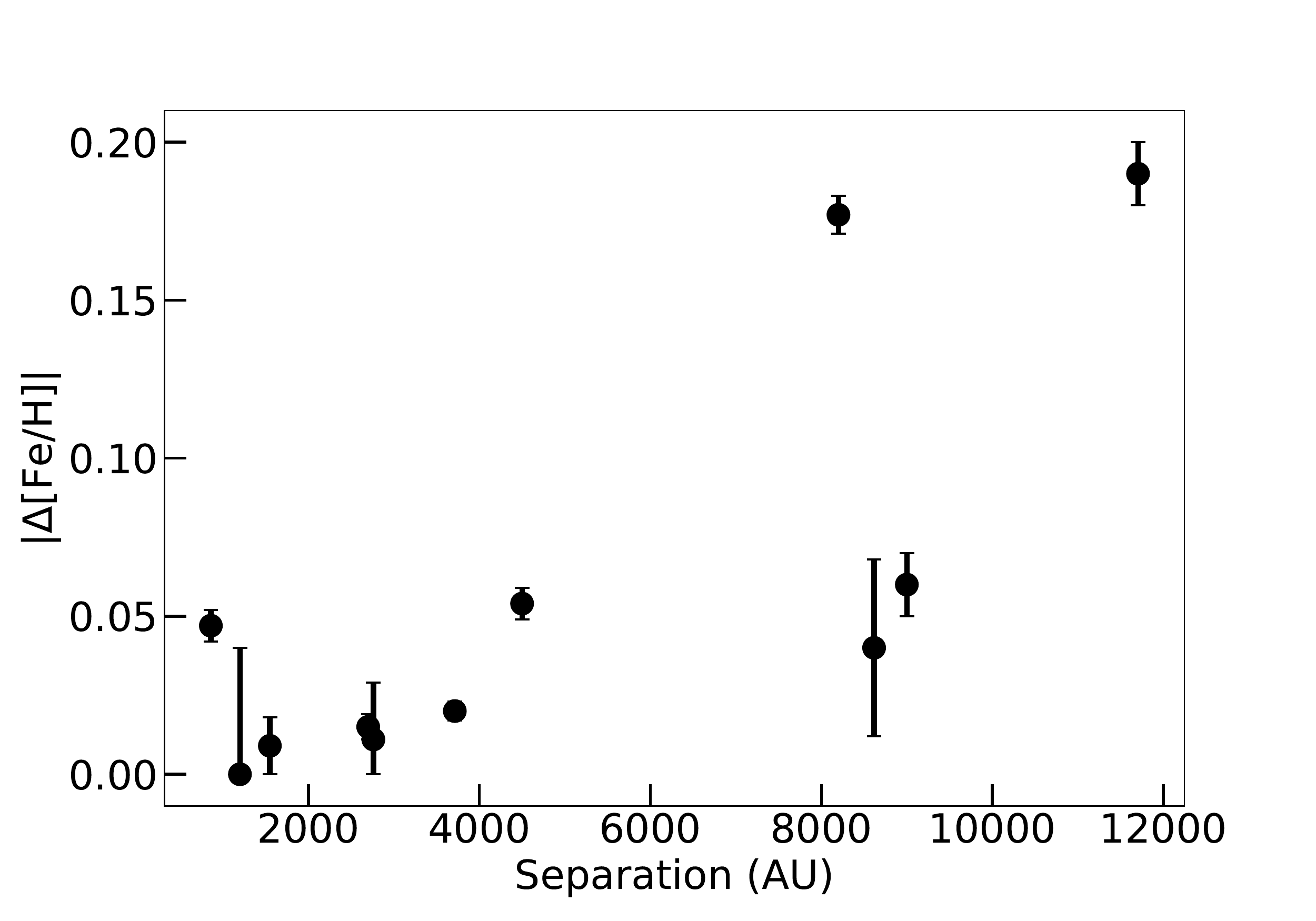}
\caption{Absolute value of iron abundance difference as a function of binary star separation for 12 twin-star binary systems with high-precision chemical abundance analysis available.}
\label{f:feh}
\end{figure}

Initially, we compared the systems' $\Delta$[Fe/H] values to their average $T_\mathrm{eff}$, $\log\,g$, and [Fe/H], as well as binary separation. $\Delta$[Fe/H] was chosen to run the first test because iron is often the most robust of all elements measured due to the large number of moderately strong and clean (unblended) iron lines in the spectra of solar-type stars. We found no significant trends when comparing $\Delta$[Fe/H] to the average effective temperature, $\log\,g$, or overall metallicity of the system. We also looked for correlations between $\Delta$[Fe/H] and the {\it differences} in stellar parameterers, finding again no significant trends. A comparison of the absolute value of $\Delta$[Fe/H] to binary separation, on the other hand, revealed a weak correlation, as shown in Figure~\ref{f:feh}. It appears that the absolute value of $\Delta$[Fe/H] increases for larger separations between the stars in a binary system. This apparent trend motivated us to compare the absolute values of all other elements to binary separation.

Although HAT-P-4 follows reasonably the weak trend shown in Figure~\ref{f:feh}, with an iron abundance difference of about 0.1\,dex and a separation of nearly 30\,000\,AU, we omitted this system from further analysis due to the fact that it has a significantly higher separation compared to the rest of the sample. The difference in separation between HAT-P-4 and the next highest separation value is over 20\,000 AU, which decreases the reliability of any existing trends for the sample. A single data point three times farther than the farthest point in Figure~\ref{f:feh} could easily drive unreliable trends.

After examining plots of absolute value of $\Delta$[X/H] versus binary separation, we noticed the same general tendency for each chemical species to increase in $|\Delta$[X/H]| as separation increased, albeit showing weak correlations in all cases, including a few marginal ones. However, we also noticed that, on average, the absolute values of $\Delta$[X/H] increased as the number associated with the absolute abundance of the chemical element in the Sun ($A_\mathrm{X}^\odot$) decreased. Since, to first approximation, elemental abundances in stars scale linearly with the solar abundances, we can use the solar abundances as a metric for which elements are more or less abundant on an absolute sense for any other star.

\begin{figure}
\includegraphics[width=\columnwidth, trim=0.6cm 1.8cm 0.5cm 0.5cm]{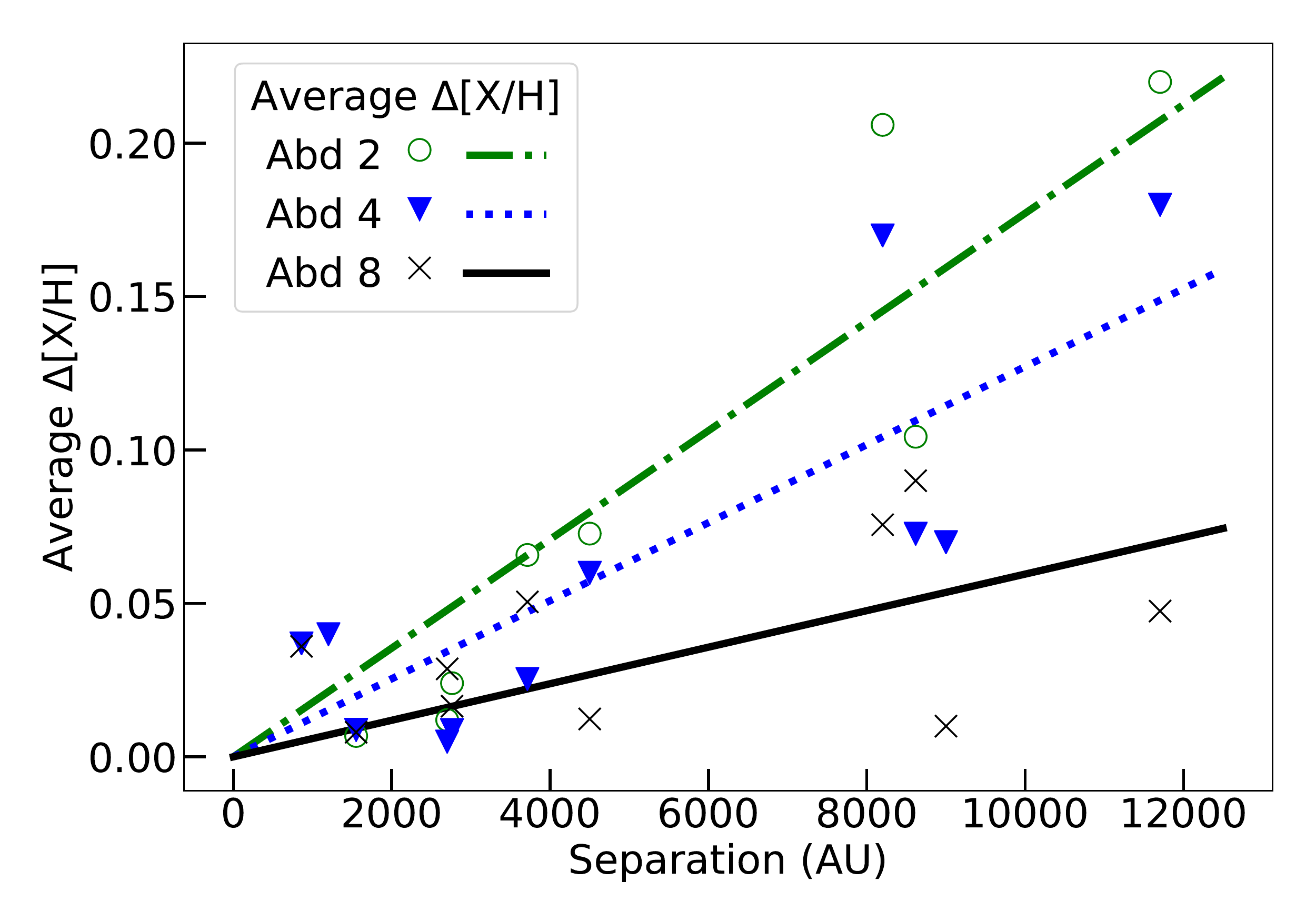}
\caption{Absolute value of average elemental abundance difference as a function of binary separation. Elements are grouped and labeled according to their absolute abundance in the Sun. For example, the ``Abd 2'' group corresponds to elements which in the Sun have an absolute abundance between $A_\mathrm{X}=2$ and $A_\mathrm{X}=3$.}
\label{f:avg}
\end{figure}

Elements with a lower absolute abundance in the Sun tend to show greater system-to-system $\Delta$[X/H] variations compared to elements that have a high solar absolute abundance. The correlation is not a very strong one, thus we grouped the chemical species from each binary system according to absolute abundance in the Sun ($A_\mathrm{X}^\odot\simeq2,3,4,\ldots,8$), and then took the average of the absolute value of the $\Delta$[X/H] values of each group in order to compare them to binary separation. The $A_\mathrm{X}^\odot\simeq2$ or ``Abd 2'' group, for example, includes all elements with absolute solar abundance between 2 and 3. An example of this process for three solar abundance groups is shown in Figure~\ref{f:avg}. Chemical species from the absolute abundance group $A_\mathrm{X}^\odot=1$ were omitted due to the scarcity of data points in that group. The elements included in each absolute solar abundance group are listed in Table~\ref{t:abundance_groups} There were only three binaries, including HAT-P-4, that had available data in that group of very low absolute solar abundance.

\begin{figure}
\includegraphics[width=\columnwidth, trim=0.2cm 1.5cm 1.5cm 1.5cm]{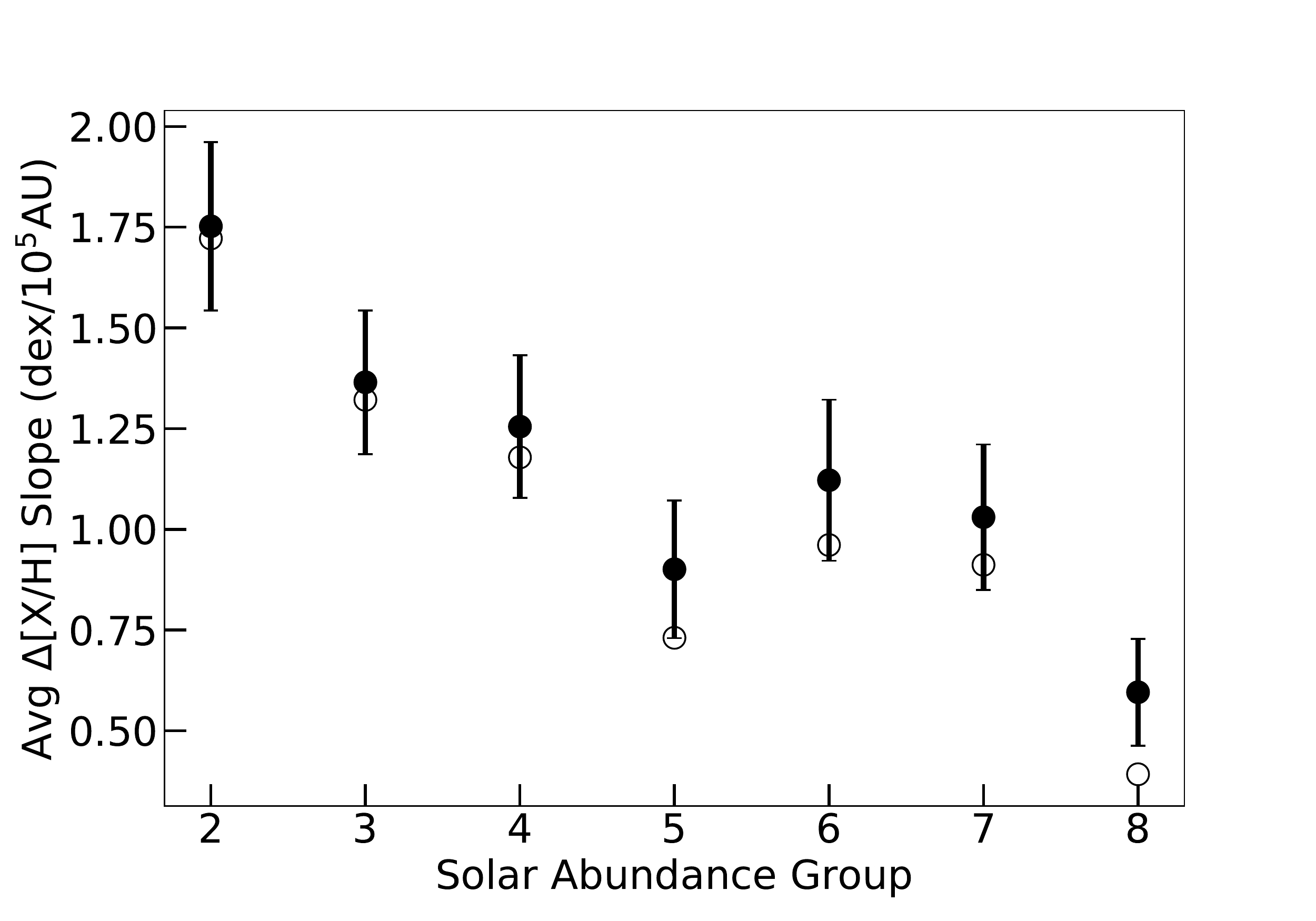}
\caption{Slope of the correlation between absolute value of average elemental abundance difference as a function of binary separation, as shown in Figure~\ref{f:avg}, versus absolute solar abundance group. The solid circles correspond to an analysis using all systems. The open circles exclude the three comoving pairs that do not satisfy the criterion of actual binarity.}
\label{f:slopes}
\end{figure}

Inspection of plots similar to those illustrated in Figure~\ref{f:avg}, but for all solar abundance groups, indicated that the slope of the average $|\Delta$[X/H]| versus separation relations decreases for groups of elements which are more abundant, on an absolute sense, in the Sun. In fact, there is a clear correlation between these slopes and the representative order-of-magnitude solar abundance for each of these element groups, as shown in Figure~\ref{f:slopes} (filled circles, which were obtained using eleven binary star candidates). The data shown in this figure as filled circles have a Pearson correlation coefficient of $-0.90$, with a p-value of 0.005, further supporting our finding of a significant anticorrelation between the strength of the abundance difference versus binary separation trend and absolute elemental abundance in the Sun.

\begin{table}
\caption{Absolute solar abundance groups.}
\label{t:abundance_groups}
\centering
\begin{tabular}{cl}
\hline
Group & Elements \\ \hline
Abd 2 & Sr, Zr, Rb, Y, Ba \\
Abd 3 & V, Sc \\
Abd 4 & Co, Ti, Zn, Cu \\ 
Abd 5 & Cr, Mn, K \\
Abd 6 & Al, Ca, Na, Ni \\
Abd 7 & N, Mg, Si, Fe, S \\
Abd 8 & O, C \\ \hline
\end{tabular}
\end{table}

If we exclude the two systems that are unbound according to Figure~\ref{f:binarity}, the trend observed in Figure~\ref{f:slopes} does not disappear. The open circles in Figure~\ref{f:slopes} correspond to that case. In fact, excluding two other systems: XO-2, which is marginally bound, and the Kronos/Krios pair, which then remains as an obvious outlier in the abundance graphs that one might think is the driver of any remaining correlations, results in an somewhat tighter downward trend, albeit with a shallower slope (this case is not shown in Figure~\ref{f:slopes}). Although the result shown in Figure~\ref{f:slopes} is based on a small number of systems, and the inclusion of some of those systems could be questionable, the main result appears to be robust. The fact that the slope values decrease when excluding systems that are more wide (and thus likely unbound today) could be due to a non-linear nature of this trend.

The result shown in Figure~\ref{f:slopes} could be attributed to a chemical inhomogeneity in the molecular clouds from which the stars in each binary formed. A trend in increasing differences in metallicity as distance increases between members of a binary system suggests that the increased distance allowed for the stars to form from increasingly different material. The differences become more apparent when the chemical element in question had a low absolute abundance to begin with. It is possible that some of the scatter observed in the abundance difference versus condensation temperature plots of twin-star binaries is due to this effect. We emphasize that our findings do not compromise the significance of these previously-mentioned $T_\mathrm{cond}$ trends and/or their interpretation as due to sequestering or ingestion of planetary material. It is possible that this trend with binary separation that we might have uncovered is simply just another piece of the puzzle.

As we did with the HIP\,34407/HIP\,34427 system before, we should briefly explore the potential impact of difussion on these results. When we looked at the difference of surface gravity in our binary stars, we found that the greatest difference we had was about 0.1, at surface gravities near 4.38 and 4.46, and the difference in [Fe/H] was 0.190 and 0.060 respectively. The same difference in surface gravity at 4.38 and 4.46 in the \cite{souto19} study shows a modeled difference in [Fe/H] of 0.02, approximately. The reason that we only chose two of our binary pairs which had the greatest difference in surface gravity for comparison to the models, is that as the difference in surface gravity in the model goes below 0.1, the difference in metallicity also decreases to values of $<0.02$. Thus, even in the most extreme cases, differential diffusion effects appear to be very small. Of course, if one of the stars in a given binary had a significantly higher internal turbulence, and thus a faster rate of diffusion than its sibling, that might affect the trend of refractories showing greater differences than volatiles. We are not completely dismissing the differential effects of turbulence in our trends of abundance difference, but from what we have seen in the studies on difusion, it does not appear to be a major contributing factor.

\section{Conclusions}

The stars in the HIP\,34407/HIP\,34426 twin-star comoving pair, arguably a true binary at formation, have clearly different chemical composition. This challenges the general assumption that stars formed from the same gas cloud have the same surface elemental abundances throughout their main-sequence lifetimes, a fundamental requirement for success in the field of chemical tagging. The abundance differences observed in this system resemble those seen in other twin-star binaries, in particular in connection to the very strong correlation between abundance difference and the elements' condensation temperature. The latter is clearly seen in all systems with any noticiable abundance differences. While there are reasons to believe that a chance alignment is unlikely, we have no conclusive proof that HIP\,34407 and HIP\,34426 formed together, a statement that could apply also to other systems with very wide separation and large elemental abundance differences. Our conclusions below are therefore subject to this important caveat.

The condensation temperature trends have been attributed to either retention of material by planets and/or engulfment of material with planet-like composition, a hypothesis that still needs further support in the form of independent confirmation from exoplanet detection surveys. Although our investigation of the HIP\,34407/HIP\,34426 system, for which high-precision (errors$\simeq0.01$\,dex) abundances are derived for the first time, provides yet another example of abundance anomalies highly correlated with $T_\mathrm{cond}$, by itself it does not allow us to conclusively establish the nature or origin of such trend. A search of planets and circumstellar material around these stars has given us interesting results thus far, but still not conclusive. Nevertheless, the same can be said for all previous investigations of chemical abundance anomalies in other twin-star binary systems.

Furthermore, for the first time we have analyzed a small sample of twin-star binary systems with high-precision chemical abundance analysis available in the literature in order to search for correlations with stellar parameters and/or binary star characteristics. Our analysis reveals a weak, but statistically significant correlation between the absolute value of elemental abundance differences and binary star separation. Interestingly, we find that species that are least abundant in the Sun's photosphere exhibit the strongest dependency on binary separation compared to those species that are more abundant. We argue that this could be an effect produced by the inhomogeneity of the gas clouds from which the binary stars of this study formed. The greater the separation, the larger the abundance differences, a trend which is anti-correlated with the absolute abundance of the element in question. Elements which are less abundant on an absolute sense will be the ones to show larger relative differences between the two stars in a binary system given that they require less of an absolute difference to become apparent.

In order to detect the correlation with binary star separation, the elemental abundance data had to be grouped together in bins of absolute solar abundance. The trend itself is derived from an analysis of more than 10 twin-star pairs. The difficulty in unveiling this trend stems from the fact that in most of these systems, the abundance anomalies are already highly correlated with the elements' condensation temperature, which is likely an independent factor from binary star separation. Our finding suggests that the $T_\mathrm{cond}$ trends are likely blurred by yet another, external effect. We propose that the latter is due to chemical inhomogeneity at the time of star and planet formation. As more twin-star binary systems are analyzed in the future at high precision (and/or more chemical elements are added to the list of existing data sets), we hope to be able to better characterize this trend with binary separation, and perhaps even disentagle it from the $T_\mathrm{cond}$ trend in order to construct a more complete model of twin-star binary abundance anomalies.

\section*{Acknowledgements}

SJL would like to thank Josie Thompson for their expertise in Python coding; programming can be as mysterious as the cosmos and Josie was there to sharpen our programming tools to accommodate our scientific needs. JC acknowledges support from CONICYT project Basal AFB-170002 and by the Chilean Ministry for the Economy, Development, and Tourism Programa Iniciativa Cient\'ifica Milenio grant IC 120009, awarded to the Millenium Institute of Astrophysics. JM thanks FAPESP (2018/04055-8).  We thank Diego Lorenzo-Oliveira and Lorenzo Spina for observing our targets with HARPS.



\bibliographystyle{mnras}
\input{hip34407-26.bbl}

\bsp	
\label{lastpage}

\end{document}